\documentclass[12pt]{article}
\pdfoutput=1
\usepackage{makeidx}
\makeindex
\usepackage[a4paper]{geometry}
\usepackage{jheppub,amsmath,amssymb,amsfonts,amsxtra, mathrsfs,graphics,graphicx,amsthm,epsfig,ytableau,bm,longtable,float,color,tikz,mathtools,xfrac,footnote,rotating,lscape}
\usepackage{textgreek}
\usetikzlibrary{decorations.pathmorphing}
\usetikzlibrary{decorations.markings}
\usetikzlibrary{arrows, decorations.markings, calc, fadings, decorations.pathreplacing, patterns, decorations.pathmorphing, positioning}
\usepackage{tikz-cd}

\usetikzlibrary{positioning,shapes}
\usetikzlibrary{chains}
\usetikzlibrary{arrows,fit,decorations.pathreplacing}
\tikzstyle{every picture}+=[remember picture]
\tikzstyle{na} = [baseline=-.5ex]

\usepackage{empheq}
\usepackage{multirow}
\usepackage{booktabs}
\usepackage[american]{babel}

\usepackage[latin1]{inputenc}

\makeatletter
\newcommand{\vast}{\bBigg@{1}}
\newcommand{\Vast}{\bBigg@{5}}
\makeatother

\usepackage{hyperref}

\numberwithin{equation}{section}

\newcommand{\ie}{\textit{i.e.}}

\numberwithin{equation}{section}

\newcommand{\nn}{\nonumber}

\newcommand{\be}{\begin{equation}} \newcommand{\ee}{\end{equation}}
\newcommand{\bea}{\begin{equation} \begin{aligned}} \newcommand{\eea}{\end{aligned} \end{equation}}

\def\U{\mathrm{U}}
\def\SO{\mathrm{SO}}

\def\SU{\mathrm{SU}}

\newcommand{\rd}{\mathrm{d}}

\DeclareMathOperator{\Tr}{Tr}

\DeclareMathOperator{\re}{\mathbb{R}e}

\newcommand{\cA}{\mathcal{A}}
\newcommand{\cB}{\mathcal{B}}

\newcommand{\cH}{\mathcal{H}}
\newcommand{\cI}{\mathcal{I}}

\newcommand{\cL}{\mathcal{L}}

\newcommand{\cN}{\mathcal{N}}
\newcommand{\cO}{\mathcal{O}}

\newcommand{\cQ}{\mathcal{Q}}

\newcommand{\cS}{\mathcal{S}}

\newcommand{\bC}{\mathbb{C}}

\newcommand{\bP}{\mathbb{P}}

\newcommand{\bR}{\mathbb{R}}

\newcommand{\fg}{\mathfrak{g}}

\usepackage[Symbol]{upgreek}
\usepackage{bm}

\usepackage{calligra}
\DeclareMathAlphabet{\mathcalligra}{T1}{calligra}{m}{n}

\title{A note on the entropy of rotating BPS $\boldsymbol{\text{AdS}_7 \times S^4}$ black holes}

\author[a,b,c]{Seyed Morteza Hosseini,}
\author[d]{Kiril Hristov}
\author[b,c]{and Alberto Zaffaroni}

\affiliation[a]{Kavli IPMU (WPI), UTIAS, The University of Tokyo, Kashiwa, Chiba 277-8583, Japan}
\affiliation[b]{Dipartimento di Fisica, Universit\`a di Milano-Bicocca, I-20126 Milano, Italy}
\affiliation[c]{INFN, sezione di Milano-Bicocca, I-20126 Milano, Italy}
\affiliation[d]{Institute for Nuclear Research and Nuclear Energy, Bulgarian Academy of Sciences, \\Tsarigradsko Chaussee 72, 1784 Sofia, Bulgaria}

\emailAdd{morteza.hosseini@ipmu.jp}
\emailAdd{khristov@inrne.bas.bg}
\emailAdd{alberto.zaffaroni@mib.infn.it}

\preprint{IPMU18-0049}
\abstract{In this note we show that the entropy of BPS, rotating, electrically charged  AdS$_7 \times S^4$ black holes can be obtained by an extremization principle involving a particular combination of anomaly coefficients of the six-dimensional $\mathcal{N} = (2, 0)$ theory.
This result extends our previous finding for BPS, rotating AdS$_5 \times S^5$ black holes.}

\begin{document}

\setcounter{tocdepth}{2}
\maketitle

%
%

\section{Introduction}
\label{sec:introduction}

There has been some recent progress in deriving the entropy of BPS static, asymptotically AdS$_4$ magnetically charged black holes
that can be embedded in string/M-theory \cite{Benini:2015eyy,Benini:2016rke,Azzurli:2017kxo,Hosseini:2017fjo,Benini:2017oxt,Bobev:2018uxk}.
The method uses a dual field theory computation based on localization. There are also many examples of BPS, electrically charged,
rotating black holes in various dimensions whose entropy cannot be yet explained in this way.
The most famous ones are asymptotic to AdS$_5\times S^5$ \cite{Gutowski:2004ez,Gutowski:2004yv,Chong:2005da,Chong:2005hr,Kunduri:2006ek}.  They depend on three electric charges  $Q_I$ $(I = 1,2,3)$, associated with rotations in $S^5$, and two angular momenta $J_\phi, J_\psi$ in AdS$_5$.
Supersymmetry actually requires a constraint among the charges and only four of them are independent.
The derivation of their entropy in terms of states of the dual ${\cal N}=4$ $\SU(N)$ super Yang-Mills (SYM) theory
is still an open problem \cite{Kinney:2005ej,Grant:2008sk,Chang:2013fba}.
The natural place where to look for such derivation is the superconformal index \cite{Romelsberger:2005eg,Kinney:2005ej} 
\bea\label{index4d}
\cI_{S^3 \times S^1} (\Delta_I , \omega_i) =
 \Tr_{\cH} (-1)^F e^{- 2 \pi i \left( \sum_{I=1}^{3} \Delta_I r_I - \sum_{i=1}^{2} \omega_i h_i \right)} \, ,
\eea
where $h_i$ are the generators of angular momentum, $r_i$ are the Cartan generators of the $\SO(6)$ R-symmetry
and $\omega_i$, $\Delta_I$ are the conjugate chemical potentials, respectively.  
\eqref{index4d} is defined for $ \Delta_1 + \Delta_2 + \Delta_3 + \omega_1 + \omega_2 \in \mathbb{Z}$
since the exponent should commute with the relevant supercharge and chemical potentials are only defined modulo one.
The index counts states preserving the same supersymmetries of the black holes and it depends on a number of fugacities
equal to the number of conserved charges of the black holes.
However, due to a large cancellation between bosonic and fermionic states, the superconformal index is a quantity of order one
for generic values of the fugacities while the entropy scales like $N^2$ \cite{Kinney:2005ej}.
We recently observed \cite{Hosseini:2017mds} that the Bekenstein-Hawking entropy of these black holes can be obtained
as the Legendre transform with respect to $\omega_i$ and $\Delta_I$ of the quantity  
\bea
  \label{modified:Casimir:SUSY:N=4}
 E^{\SU(N)}= - i \pi  N^2  \frac{\Delta_1 \Delta_2 \Delta_3}{\omega_1 \omega_2} \, ,
\eea
with the determination
\bea
 \label{constraint:N=4}
 \sum_{I=1}^{3} \Delta_I + \sum_{i=1}^{2} \omega_i = 1 \, .
\eea
This constraint among chemical potentials reflects the constraint among charges of the black holes
and is compatible with the constraint in the index \eqref{index4d}.
The quantity \eqref{modified:Casimir:SUSY:N=4} can be expressed in terms of the flavored cubic t'Hooft anomaly coefficients
of ${\cal N}=4$ SYM. Indeed it can be obtained by an equivariant integral of the anomaly polynomial of the theory, as shown  in \cite{Bobev:2015kza}.%
\footnote{With a different choice of angular determination for the chemical potentials $\Delta_I$ and $\omega_i$, the quantity \eqref{modified:Casimir:SUSY:N=4} becomes the supersymmetric Casimir energy of the theory (see for example Eq.\,(4.50) in \cite{Bobev:2015kza}). The supersymmetric Casimir energy can be interpreted as the energy of the vacuum \cite{Assel:2015nca}
and it arises both as a prefactor in the relation that connects the supersymmetric partition function on $S^3\times S^1$
to the index \cite{Assel:2014paa,Lorenzen:2014pna} and also as a term in the high-temperature expansion of the index \cite{Ardehali:2015hya,Ardehali:2015bla}.
All these results seem to have been obtained assuming  a particular angular determination of fugacities, which implies, in particular,  
$\Delta_1 + \Delta_2 + \Delta_3 + \omega_1 + \omega_2 = 0$, instead of \eqref{constraint:N=4}.}

In this short note we extend our observation to BPS, electrically charged, rotating black holes in AdS$_7\times S^4$.
We expect a family of black holes depending on two electric charges $Q_I$ $(I = 1,2)$, associated with rotations in $S^4$,
and three angular momenta $J_i$ $(i=1,2,3)$ in AdS$_5$. Supersymmetry again requires a constraint among the charges
and only four of them are independent. The dual field theory is the $A_{N-1}$ $\cN = (2,0)$ theory in six dimensions.
Inspired by the AdS$_5$ result, we consider the expression for the equivariant integral of the anomaly polynomial of the theory, which,  at large $N$, is given by \cite{Bobev:2015kza}
\be\label{6dCasimirenergy}
 E^{(A_{N-1})} (\Delta_I , \omega_i) = i \pi N^3 \frac{\left( \Delta_1 \Delta_2 \right)^2}{12 \omega_1 \omega_2 \omega_3} \, .
\ee
We will show that the Bekenstein-Hawking entropy of seven-dimensional BPS black holes can be obtained by extremizing the quantity
\be\label{extremiz}
 - E^{(A_{N-1})} (\Delta_I , \omega_i)
 - 2 \pi i \sum_{I=1}^{2} \Delta_I Q_I - 2 \pi i \sum_{i=1}^{3} \omega_i J_i \, ,
\ee
with respect to $\Delta_I$, $\omega_i$ with the constraint
\be\label{constraint6d}
 \sum_{I=1}^{2} \Delta_I + \sum_{i=1}^{3} \omega_i = 1 \, .
\ee
The general black hole solution depending on all four conserved quantities is not available
but we will check that \eqref{extremiz} correctly reproduces the entropy of the existing solutions.
A two-parameter black hole, with two electric charges and one angular momentum, was found in \cite{Cvetic:2005zi}
as the BPS limit of a non-extremal solution \cite{Chong:2004dy}.
The solution was later extended to a three-parameter one, with three rotations and one electric charge, in \cite{Chow:2007ts}.
We have explicitly checked the validity of \eqref{extremiz} in both cases
and are thus confident that the result holds in general.

It is remarkable that the equivariant integral of the anomaly polynomial accounts for the entropy of both the AdS$_5\times S^5$ and  AdS$_7\times S^4$ supersymmetric black holes.
Moreover, it is  noteworthy  that,  in both cases, the solutions to the extremization equations associated with \eqref{modified:Casimir:SUSY:N=4} and \eqref{6dCasimirenergy}  are complex but the value of the Legendre transform at the critical points, the entropy,  is a {\it real} function of the black hole charges.  This result still needs a proper field theory interpretation.
Here we just make few observations.

According to the standard holographic dictionary, the black hole entropy should account for the $1/16-$BPS states in the $\cN = (2,0)$
theory with given electric charge and angular momentum. As in four dimensions, the partition function for such BPS states is too difficult to compute
due to the small amount of preserved supersymmetry. The superconformal index
of the $\cN = (2,0)$ theory, on the other hand,  counts states weighted with signs,\footnote{In our notations, $Q_I$ are eigenstates of $r_I$ and $J_i$ of $-h_i$.} 
\be\label{index6d}
 \cI_{S^5 \times S^1} (\Delta_I , \omega_i) =
 \Tr_{\cH} (-1)^F  e^{- 2 \pi i  \left( \sum_{I=1}^{2} \Delta_I r_I - \sum_{i=1}^{3} \omega_i h_i \right)} \, ,
\ee
where $h_i$ are the generators of angular momentum, $r_I$ are the Cartan generators of the $\SO(5)$ R-symmetry,
and $ \Delta_1 + \Delta_2  + \omega_1 + \omega_2 + \omega_3\in \mathbb{Z}$.
The index is explicitly computable but, as in four dimensions, is a quantity of order one for generic values of the fugacities.
The $S^5 \times S^1$ partition function, on the other hand, is related to the index in the large $N$ limit by \cite{Kim:2012ava}  
\be
 \log Z_{S^5 \times S^1} (\Delta_I , \omega_i) = - E_{\text{susy}} (\Delta_I , \omega_i)
 + \log \cI_{S^5 \times S^1} (\Delta_I , \omega_i) \, ,
\ee
where $E_{\text{susy}}$ is the supersymmetric Casimir energy and scales like $\cO(N^3)$. 
It would be interesting to see if the quantity \eqref{6dCasimirenergy} dominates the index or the partition function
in some particular regime for the chemical potentials and a choice of angular determination compatible with the constraint \eqref{constraint6d}.
An expression for $E_{\text{susy}}$ has been conjectured in \cite{Bobev:2015kza} by integrating the anomaly polynomial of the $\cN = (2,0)$ theory,
and formally coincides with \eqref{extremiz}. However, the conjecture seems to assume a different angular determination for the chemical potentials, 
compatible with  $\sum_{I=1}^{2} \Delta_I + \sum_{i=1}^{3} \omega_i = 0$ rather than  \eqref{constraint6d}. 

The AdS$_5\times S^5$ and AdS$_7\times S^4$  black holes behave quite differently from their magnetically charged relatives
in AdS$_4\times S^7$ whose entropy has been recently derived \cite{Benini:2015eyy,Benini:2016rke}.
The main difference comes from the magnetic charges that have a dual interpretation as a topological twist \cite{Witten:1988ze,Witten:1991zz}.
The topologically twisted index \cite{Benini:2015noa,Nekrasov:2014xaa} has been shown to scale like $N^{3/2}$
in the large $N$ limit \cite{Benini:2015eyy,Hosseini:2016tor,Hosseini:2016ume}, suggesting that there is no cancellation
between bosonic and fermionic ground states, while the superconformal index is a quantity of order one.
It should also be noticed that the derivation of the entropy of AdS$_4\times S^7$ black holes is a purely microscopic counting
with no reference to the ubiquitous Cardy formula \cite{Cardy:1986ie}. On the other hand, whatever its field theory interpretation is,
the extremization principles in five and seven dimensions suggests that some sort of Cardy mechanism is at work.
First of all, as already said, \eqref{modified:Casimir:SUSY:N=4} and \eqref{6dCasimirenergy} can be written in terms of
anomaly coefficients for the R and flavor symmetries of the dual theory. This is simple to see for \eqref{modified:Casimir:SUSY:N=4}.
Indeed, by an obvious redefinition of the chemical potentials (see Appendix \ref{app:casimir}), \eqref{modified:Casimir:SUSY:N=4} can be written as the large $N$ limit of%
\footnote{Comparing to expressions for the supersymmetric Casimir energy  that can be found in
\cite{Assel:2014paa,Lorenzen:2014pna,Assel:2015nca} we have an extra minus one in the numerator which is due to the constraint \eqref{constraint6d}.}
\be
 E^{\SU(N)} = \frac{4 \pi i}{27}  \frac{(\omega_1+\omega_2-1)^3}{\omega_1\omega_2}a(\hat \Delta_I) \, .
\ee
Here 
\be
 a(\hat \Delta_I)=  \frac{9}{32} \Tr R^3(\hat \Delta_I)
 = \frac{27}{32}(N^2-1) \hat\Delta_1 \hat \Delta_2 \hat \Delta_3 \, ,
\ee
together with $\hat \Delta_1 +\hat\Delta_2+\hat\Delta_3= 2$ is the trial central charge of ${\cal N}=4$ SYM.
\eqref{6dCasimirenergy} can be written similarly since it arises from an equivariant integration of the eight-form anomaly polynomial of the $6{\rm D}$ $\cN = (2,0)$ theory \cite{Bobev:2015kza}. Moreover, for an extremal BTZ black hole in AdS$_3$ the relevant quantity to consider is the elliptic genus,
whose logarithm in the large $N$ limit goes as $c_l / \omega$ where $\omega$ is the chemical potential associated with rotation
and $c_l$ is the left-moving central charge. The entropy of a black hole of angular momentum $j$ is then correctly reproduced
by the Legendre transform with respect to $\omega$, \ie\,$S \propto \sqrt{j c_l}$. Obviously, this is nothing else than Cardy formula.

The paper is organized as follows. In section \ref{BH} we first review the basic features of the relevant truncation of 
seven-dimensional maximal gauged supergravity and we later discuss the BPS, rotating black holes of interest.
In section \ref{sec:entropy}, we show that the Bekenstein-Hawking entropy of
the black holes can be obtained as the Legendre transform of the quantity \eqref{6dCasimirenergy}.
We conclude in section \ref{sec:discussion} with discussions and future directions.
In the appendices  we briefly discuss a conjecture to extend
our previous result for black holes in AdS$_5\times S^5$ to more general  compactifications and the dimensional reduction of the seven-dimensional  black holes to six dimensions.

\section{Supersymmetric rotating AdS$_7$ black holes}\label{BH} 

The supersymmetric rotating AdS$_7$ black holes of interest are solutions of the $\SO(5)$ maximal ($\cN = 4$) gauged supergravity
in seven dimensions \cite{Pernici:1984xx}, obtained by reducing eleven-dimensional supergravity on $S^4$ \cite{Nastase:1999cb,Nastase:1999kf}.
We will work with a $\U(1)^2$ consistent truncation \cite{Liu:1999ai}
of the theory, which consists of the metric,
a three-form potential $A_{(3)}$, two Abelian gauge fields $A_{(1)}^I$ $(I=1,2)$ in the Cartan of $\SO(5)$
and two real scalars $\varphi_1$ and $\varphi_2$.
The bosonic Lagrangian is given by \cite{Pernici:1984xx}%
\footnote{Here we use the conventions of \cite{Cvetic:2005zi}.}
\bea
 \label{7d:Lag}
 \cL_7 & = R\, \star 1 - \frac12 \sum_{i=1}^2 {\star \rd \varphi_i} \wedge \rd \varphi_i -
 \frac12 \sum_{I=1}^2 L_I^{-2}\, {\star F_{(2)}^I} \wedge F_{(2)}^I
 - \frac12 (L_1 L_2)^2 {\star F_{(4)}} \wedge F_{(4)} \\
 & - 2 g^2 \left[ (L_1 L_2)^{-4} - 8 L_1 L_2 - 4 L_1^{-1} L_2^{-2} - 4 L_1^{-2} L_2^{-1} \right] \star 1 & \\
 & - g F_{(4)} \wedge A_{(3)}
 + F_{(2)}^1 \wedge F_{(2)}^2 \wedge A_{(3)} \, ,
\eea
where
\bea
 F_{(2)}^I & = \rd A_{(1)}^I \, , \qquad \qquad \qquad && F_{(4)} = \rd A_{(3)} \, , \\
 L_1 & = e^{-\frac1{\sqrt2} \varphi_1 -\frac1{\sqrt{10}} \varphi_2} \, ,
 && L_2 = e^{\frac1{\sqrt2} \varphi_1 - \frac1{\sqrt{10}} \varphi_2} \, ,
\eea
and $g$ is the gauge coupling constant.
There is a ``first-order self-duality'' condition for the four-form field strength
that has to be imposed after the variation of the Lagrangian and that
can be conveniently written by including a two-form potential $A_{(2)}$, and defining
\be
F_{(3)} = \rd A_{(2)} - \frac12 A_{(1)}^1 \wedge \rd A_{(1)}^2 - \frac12 A_{(1)}^2 \wedge \rd A_{(1)}^1 \, .
\ee
The self-duality equation then reads
\be
 (L_1 L_2)^2 {\star F_{(4)}} = -2 g A_{(3)} - F_{(3)} \, .
\ee
We are interested in supersymmetric black holes with electric charges $Q_I$ $(I = 1,2)$ under the $\U(1)^2$ and
angular momenta $J_i$ $(i=1,2,3)$ in AdS$_7$. We expect supersymmetry to impose a constraint among the five charges,
leaving four independent ones. The most general family of such black holes has not been written yet.
A two-parameter black hole, with two electric charges and one angular momentum with a constraint among them,
was found in \cite{Cvetic:2005zi}. A three-parameter family of black holes,
with three rotations and one electric charge with a constraint, was later found in  \cite{Chow:2007ts}. Note that in both cases the near-horizon geometry is a warped product of AdS$_2$ and a squashed $S^5$. We now write explicitly these solutions and their thermodynamic quantities. We have corrected few misprints in \cite{Cvetic:2005zi}.

\subsection{Single-rotation two-charge black holes}
\label{sec:single-rotation}

The solution can be written as \cite{Cvetic:2005zi}
\bea
 \label{7d:bh:CGLP:metric}
 \rd s^2 & = (H_1 H_2)^{1/5} \Big(- \frac{V} {H_1 H_2 B} r^2 \rd t^2
 + B (\sigma + f \rd t)^2 + \frac{\rd r^2}{V} + r^2 \rd s^2_{\bC\bP^2} \Big) \, , \\
 A_{(1)}^I & = \frac{2 m s_I}{\rho^4 \Xi H_i} (\alpha_I\ \Xi_- \rd t + \beta_I \sigma) \, , \\
 A_{(2)} & = \frac{m a s_1 s_2}{\rho^4 \Xi_-} \left( \frac{1}{H_1} + \frac{1}{H_2} \right) \rd t \wedge \sigma \, ,\quad
 A_{(3)} = \frac{2 m a s_1 s_2}{\rho^2 \Xi \Xi_-} \sigma \wedge J \, , \\
 L_I & = (H_1 H_2)^{2/5} H_I^{-1} \, , \quad
 H_I = 1 + \frac{2 m s_I^2}{\rho^4} \, , \quad
 \rho = \sqrt{\Xi} r \, , \\
 \alpha_1 & = c_1 - \frac12 (1 - \Xi_+^2) (c_1 - c_2) \, , \quad
 \alpha_2 = c_2 + \frac12 (1 - \Xi_+^2) (c_1 - c_2) \, , \\
 \beta_1 & = - a \alpha_2 \, , \quad
 \beta_2 = - a \alpha_1 \, , \quad
 \Xi_{\pm} = 1 \pm a g \, , \quad
 \Xi = 1 - a^2 g^2 \, , \\
 s_I & \equiv \sinh \delta_I \, , \quad c_I \equiv \cosh \delta_I \, .
\eea
The metric functions $V$, $B$ and $f$ depends on the radial coordinate $r$ and are given by
\bea
 V & = \frac{Y}{\Xi \rho^6} \, , \quad
 B = \frac{f_1}{H_1 H_2 \Xi^2 \rho^4} \, , \quad
 f = - \frac{2 f_2 \Xi_-}{f_1} \, ,
\eea
where
\bea
 f_1 & = \Xi \rho^6 H_1 H_2 - \frac{4 \Xi_+^2 m^2 a^2 s_1^2 s_2^2}{\rho^4}
 + \frac12 m a^2 \left[ 4 \Xi_+^2 - 2 c_1 c_2 (\Xi_+^4 - 1) + (c_1^2 + c_2^2) (\Xi_+^2 - 1)^2 \right] \, , \\
 f_2 & = - \frac12 g \Xi_+ \rho^6 H_1 H_2
 + \frac14 m a \left[ 2 c_1 c_2 (\Xi_+^4 + 1) - (c_1^2 + c_2^2) (\Xi_+^4 - 1) \right] \, , \\
 Y & = g^2 \rho^8 H_1 H_2 + \Xi \rho^6
 + \frac12 m a^2 \left[ 4 \Xi_+^2 - 2 c_1 c_2 (\Xi_+^4 - 1) + (c_1^2 + c_2^2) (\Xi_+^2 - 1)^2 \right] \\
 & - \frac12 m \rho^2 \left[ 4 \Xi + 2 c_1 c_2 a^2 g^2 (3 a^2 g^2 + 8 a g + 6)
 - (c_1^2 + c_2^2) a^2 g^2 (a g + 2) (3 a g + 2) \right] \, .
\eea
Only two parameters are independent due to the constraints%
\footnote{We correct a misprint in \cite{Cvetic:2005zi} here, \ie\,$m_{\text{here}} = (3 e^{\delta_1 + \delta_2} - 1) m_{\text{there}}$.}
\bea
 & e^{\delta_1 + \delta_2} = 1 - \frac{2}{3 a g} \, , \\
 & m = \frac{128 e^{\delta_1 + \delta_2} (3 e^{\delta_1 + \delta_2} - 1)^3}
 {729 g^4 (e^{2 \delta_1} - 1) (e^{2 \delta_2} - 1) (e^{\delta_1 + \delta_2} + 1)^2 (e^{\delta_1 + \delta_2} - 1)^4} \, .
\eea
The former comes from the BPS condition and the latter is required in order to avoid naked closed timelike curves (CTCs).
With these choices, the function $V$ becomes
\bea
 V & = \frac{g^2 (r^2 - r_0^2)^2}{r^2} \left( 1 + \frac{9 e^{2( \delta_1 + \delta_2 )} - 6 e^{\delta_1 + \delta_2} + 17}
 {3 (e^{\delta_1 + \delta_2} + 1) (3 e^{\delta_1 + \delta_2} - 5) g^2 r^2} + \frac{h}{g^4 r^4} \right) \, ,
\eea
where
\bea
 h = & \Big[ 32 \big( - 2 d_1^2 - 2 d_2^2 + 9 d_1 d_2 + 9 d_1^5 d_2^5 - 3 d_1^3 d_2^3 (d_1 + d_2)^2
 + 2 d_1^2 d_2^2 (2 d_1^2 - 3 d_1 d_2 + 2 d_2^2) \\
 & - d_1 d_2 (3 d_1^2 - 2 d_1 d_2 + 3 d_2^2) \big) \Big] \bigg/
 \Big[ 9 d_1 d_2 (d_1^2 - 1) (d_2^2 - 1) (d_1 d_2 + 1) (3 d_1 d_2 - 5)^2 \Big] \, ,
\eea
and we defined $d_I \equiv e^{\delta_I}$ $(I = 1 , 2)$.
The black hole has an event horizon at $V(r_0) = 0$ which reads
\be
 r_0^2 = \frac{16}{3 g^2 (e^{\delta_1 + \delta_2} + 1) (3 e^{\delta_1 + \delta_2} - 5)} \, .
\ee
The line element $\rd s^2_{\bC\bP^2}$ in \eqref{7d:bh:CGLP:metric} is the standard Fubini-Study metric on $\bC\bP^2$:
\bea
 \rd s^2_{\bC\bP^2} = \rd \xi^2 + \frac14 \sin^2 \xi \left( \sigma_1^2 + \sigma_2^2 + \cos^2 \xi \, \sigma_3^2 \right) \, ,
\eea
where $\sigma_i$ $(i=1,2,3)$ are left-invariant one-forms on $\SU(2)$,
satisfying $\rd \sigma_i = - \frac12 \epsilon_{ijk} \sigma_j \wedge \sigma_k$.
Note that, the K\"ahler form on $\bC\bP^2$ is $J = \frac12 \rd \cB$ with $\cB = \frac12 \sin^2 \xi \sigma_3$
being the connection of the $\U(1)$ bundle over $\bC\bP^2$ whose total space is the unit $S^5$.
We also have $\sigma = \rd \psi + \cB$ and $0 \leq \psi \leq 2 \pi$ is the coordinate along the $\U(1)$ fiber of $S^5$.
The thermodynamic quantities are given by%
\footnote{We correct a misprint in \cite{Cvetic:2005zi} here, \ie\,$S_{\text{here}} = \frac14 S_{\text{there}}$.}
\bea
 E & = \frac{m \pi^2}{32 G_{\text{N}} \Xi^4} \Big[ 12 \Xi_+^2 (\Xi_+^2 - 2) - 2 c_1 c_2 a^2 g^2
 (21 \Xi_+^4 -  20 \Xi_+^3 - 15 \Xi_+^2 - 10 \Xi_+ - 6) \\
 & + (c_1^2 + c_2^2) (21 \Xi_+^6 -  62 \Xi_+^5 + 40 \Xi_+^4 + 13 \Xi_+^2 - 2 \Xi_+ + 6) \Big] \, , \\
 S & = \frac{\pi^3}{4 G_{\text{N}}} [B (r_0) H_1 (r_0) H_2 (r_0) ]^{1/2}r_0^4 \, , \\
 J & = \frac{m a \pi^2}{16 G_{\text{N}} \Xi^4} \Big[ 4 a g \Xi_+^2 - 2 c_1 c_2 (2 \Xi_+^5 - 3 \Xi_+^4 - 1)
 + a g (c_1^2 + c_2^2) (\Xi_+ +  1) (2 \Xi_+^3 - 3 \Xi_+^2 - 1) \Big] \, , \\
 Q_I & = \frac{m \pi^2 s_I}{4 G_{\text{N}} \Xi^3} \left[ a^2 g^2 \frac{c_1 c_2}{c_I} (2 \Xi_+ + 1)
 - c_I ( 2 \Xi_+^3 - 3 \Xi_+^2 - 1 ) \right] \, ,
 \\ T & = 0 \, , \qquad \Omega = - g \, , \qquad \Phi_I = -1 \, .
\eea
The charges satisfy the BPS condition $E + 3 g J - \sum_{I = 1}^{2} Q_I = 0$.

\subsection{Three-rotation single-charge black holes}

The solution reads \cite{Chow:2007ts}
\bea
 \label{7d:bh:Chow:metric}
 \rd s^2 = & H^{2/5} \bigg\{\frac{(r^2 + y^2)(r^2 + z^2)}{R} \rd r^2
 + \frac{(r^2 + y^2)(y^2 - z^2)}{Y} \rd y^2
 + \frac{(r^2 + y^2)(z^2 - y^2)}{Z} \rd z^2 \\
 & - \frac{R}{H^2 (r^2 + y^2)(r^2 + z^2)} \cA^2 \\
 & + \frac{Y}{(r^2 + y^2)(y^2 - z^2)}
 \left[ \rd t + (z^2 - r^2) \rd \psi_1 - r^2 z^2 \rd \psi_2 - \frac{q}{H (r^2 + y^2)(r^2 + z^2)} \cA \right]^2 \\
 & + \frac{Z}{(r^2 + y^2)(z^2 - y^2)}
 \left[ \rd t + (y^2 - r^2) \rd \psi_1 - r^2 y^2 \rd \psi_2 - \frac{q}{H (r^2 + y^2)(r^2 + z^2)} \cA \right]^2 \\
 & + \frac{a_1^2 a_2^2 a_3^2}{r^2 y^2 z^2} \bigg[ \rd t + (y^2 + z^2 - r^2) \rd \psi_1 +
 (y^2 z^2 - r^2 y^2 - r^2 z^2) \rd \psi_2 - r^2 y^2 z^2 \rd \psi_3 \\
 & - \frac{q}{H (r^2 + y^2)(r^2 + z^2)} \left( 1 + \frac{g y^2 z^2}{a_1 a_2 a_3} \right) \cA \bigg]^2 \bigg\} \, ,
\eea
\bea
 & L = H^{-1/5} \, , \qquad A_{(1)} = - \frac{q (1 - a_1 g - a_2 g - a_3 g)}{H (r^2 + y^2)(r^2 + z^2)} \cA \, , \\
 & A_{(2)} = \frac{q}{H (r^2 + y^2) (r^2 + z^2)} \cA \wedge \\
 & \bigg\{ \rd t + \sum_{i=1}^{3} a_i^2 ( g^2 \rd t + \rd \psi_1 )
 + \sum_{i<j} a_i^2 a_j^2 ( g^2 \rd \psi_1 + \rd \psi_2 )
 + a_1^2 a_2^2 a_3^2 ( g^2 \rd \psi_2 + \rd \psi_3 ) \\
 & - g^2 (y^2 + z^2) \rd t - g^2 y^2 z^2 \rd \psi_1
 + a_1 a_2 a_3 g \left[ \rd \psi_1 + (y^2 + z^2) \rd \psi_2 + y^2 z^2 \rd \psi_3 \right] \bigg\} \nn \, , \\
 & A_{(3)} = q a_1 a_2 a_3 \left[ \rd \psi_1 + (y^2 + z^2) \rd \psi_2 + y^2 z^2 \rd \psi_3 \right] \\
 & \wedge \left[ \frac{1}{(r^2 + y^2) z} \rd z \wedge \left( \rd \psi_1 + y^2 \rd \psi_2 \right)
 + \frac{1}{(r^2 + z^2) y} \rd y \wedge \left( \rd \psi_1 + z^2 \rd \psi_2 \right) \right] \\
 & - q g \cA \wedge \left[ \frac{z}{r^2 + y^2} \rd z \wedge \left( \rd \psi_1 + y^2 \rd \psi_2 \right)
 + \frac{y}{r^2 + z^2} \rd y \wedge \left( \rd \psi_1 + z^2 \rd \psi_2 \right) \right] \, ,
\eea
where
\bea
 R & = \frac{(r^2 - r_0^2)^2}{r^2} \left\{ g^2 r^4 + \left[ 1 + (a_1^2 + a_2^2 + a_3^2) g^2 + 2 g^2 r_0^2 \right] r^2
 + \frac{(a_1 a_2 a_3 - q g)^2}{r_0^4} \right\} \, ,\\
 Y & = \frac{1 - g^2 y^2}{y^2} \prod_{i=1}^{3} \left( a_i^2 - y^2 \right) \, , \qquad
 Z = \frac{1 - g^2 z^2}{z^2} \prod_{i=1}^{3} \left( a_i^2 - z^2 \right) \, , \\
 \cA & = \rd t + (y^2 + z^2) \rd \psi_1 + y^2 z^2 \rd \psi_2 \, , \\
 H & = 1 + \frac{q}{(r^2 + y^2)(r^2 + z^2)} \, .
\eea
The black hole has an event horizon at $r = r_0$:
\be
 r_0^2 = \frac{a_1 a_2 + a_2 a_3 + a_3 a_1 - a_1 a_2 a_3 g}{1 - a_1 g - a_2 g - a_3 g} \, .
\ee
We also denote $\Xi_i = 1 - a_i^2 g^2$ and $\Xi_{i\pm} = 1 \pm a_i g$, which are positive to have a correct signature.
The parameters $q$ and $a_i$ have to satisfy
\be
 q = - \frac{\Xi_{1-} \Xi_{2-} \Xi_{3-} (a_1 + a_2) (a_2 + a_3) (a_1 + a_3)}{(1 - a_1 g - a_2 g - a_3 g)^2 g} \, ,
\ee
in order for the solution to be free from naked CTCs.
The thermodynamic quantities are given by
\bea
 E & = - \frac{\pi^2}{8 G_{\text{N}}} \frac{\prod_{k < l} (a_k + a_l)
 \left[ \sum_{i} \Xi_i + \sum_{i < j} \Xi_i \Xi_j - \left(1 + a_1 a_2 a_3 g^3\right)
 \left( 2 + \sum_{i} a_i g + \sum_{i < j} a_i a_j g^2 \right) \right]}
 {\Xi_{1+} \Xi_{2+} \Xi_{3+} (1 - a_1 g - a_2 g - a_3 g)^2 g r_0} \, , \\
 S & = - \frac{\pi^3}{4 G_{\text{N}}} \frac{(a_1 + a_2) (a_2 + a_3) (a_1 + a_3)
 \left( a_1 a_2 + a_2 a_3 + a_1 a_3 - a_1 a_2 a_3 g \right)}
 {\Xi_{1+} \Xi_{2+} \Xi_{3+} (1 - a_1 g - a_2 g - a_3 g)^2 g r_0} \, , \\
 J_i & = - \frac{\pi^2}{8 G_{\text{N}}} \frac{(a_1 + a_2) (a_2 + a_3) (a_1 + a_3)
 \left[a_i - (a_i^2 + 2 a_i \sum_{j \neq i} a_j + \prod_{j \neq i} a_j) g + a_1 a_2 a_3 g^2 \right]}
 {\Xi_{1+} \Xi_{2+} \Xi_{3+} \Xi_{i+} (1 - a_1 g - a_2 g - a_3 g)^2 g} \, , \\
 Q & = - \frac{\pi^2}{4 G_{\text{N}}} \frac{(a_1 + a_2) (a_2 + a_3) (a_1 + a_3)}
 {\Xi_{1+} \Xi_{2+} \Xi_{3+} (1 - a_1 g - a_2 g - a_3 g) g} \, , \\
 T & = 0 \, , \qquad \Omega_i = - g \, , \qquad \Phi = -1 \, .
\eea
Finally, the charges satisfy the BPS condition
\be
 E + g \sum_{i=1}^{3} J_i - 2 Q = 0 \, .
\ee

\section{An extremization principle for the entropy}\label{sec:entropy}

In this section we will show that the Bekenstein-Hawking entropy of the BPS black holes \eqref{7d:bh:CGLP:metric} and \eqref{7d:bh:Chow:metric} can be obtained as a Legendre transform of a combination of anomaly coefficients of the dual $\cN = (2,0)$ theory in six dimensions.
The result is a natural generalization of the analogous one for AdS$_5\times S^5$ black holes \cite{Hosseini:2017mds}.

\subsection{Anomaly polynomials for $6 {\rm D}$ $\cN=(2,0)$ field theories}\label{sec:Casimir}

Our analysis involves a quantity formally equal to the supersymmetric Casimir energy of the theory, which we now briefly review.

The supersymmetric Casimir energy, $E^{(\fg)} $, for an $\cN = (2,0)$ theory with algebra $\fg$ arises
in the regularization of the $S^5 \times S^1$ partition function \cite{Kim:2012ava,Kallen:2012zn,Kim:2012tr,Lockhart:2012vp,Kim:2012qf,Minahan:2013jwa,Kim:2013nva}
and is related to the superconformal index defined in \eqref{index6d} by
\be\label{partitionf}
 \log Z_{S^5 \times S^1} (\Delta_I , \omega_i) = - E^{(\fg)} (\Delta_I , \omega_i)
 + \log \cI_{S^5 \times S^1} (\Delta_I , \omega_i) \, .
\ee
It is the leading contribution to $\log Z_{S^5 \times S^1}$ for $\beta\rightarrow\infty$, where $\beta$
is the radius of $S^1$ when the chemical potentials are rescaled as $\Delta_I=\beta \hat \Delta_I$ and $\omega_i=\beta \hat \omega_i$.%
\footnote{We have reabsorbed a standard factor of $\beta$ in the definition of  $E^{(\fg)}$ for convenience.
\eqref{partitionf} is usually written as $\log Z_{S^5 \times S^1} = -\beta E^{(\fg)} + \log \cI_{S^5 \times S^1} $.}
Since the  superconformal index is a quantity of order one for generic values of the fugacities,
the supersymmetric Casimir energy, which scales as $N^3$, is also the leading contribution to the $S^5 \times S^1$ partition function
in the large $N$ limit. 
The supersymmetric Casimir energy of the $\cN = (2,0)$ theory has been conjectured
to be equal to an equivariant integral of the eight-form anomaly polynomial and it reads \cite{Bobev:2015kza}
\be\label{equivariant}
 E^{(\fg)} (\Delta_I , \omega_i) = r_\fg E^{(1)} (\Delta_I , \omega_i)
 + \frac{i \pi}{12} d_\fg h_\fg^\lor \frac{\left( \Delta_1 \Delta_2 \right)^2}{\omega_1 \omega_2 \omega_3} \, ,
\ee
where $r_\fg$, $d_\fg$ and $h_\fg^\lor$ are the rank, dimension and dual Coxeter number
of the simply laced Lie algebra $\fg$, respectively; $E^{(1)}$ is the supersymmetric Casimir energy
of the Abelian tensor multiplet theory:
\be
 E^{(1)} (\Delta_I , \omega_i) = \frac{i \pi}{24 \omega_1 \omega_2 \omega_3}
 \left[ \left( \Delta_1 \Delta_2 \right)^2 - \sum_{i<j} \left( \omega_i \omega_j \right)^2
 + \frac14 \left( \sum_{i=1}^3 \omega_i^2 - \Delta_1^2 - \Delta_2^2 \right) \right] \, .
\ee
Here $\Delta_I$ $(I=1,2)$ are the chemical potentials conjugate to the R-symmetry generators $r_I$
and $\omega_i$ $(i=1,2,3)$ are the chemical potentials conjugate to the Cartan generators of rotations $h_i$
in three orthogonal planes in $\bR^6$. 

Superconformal indices are  defined in general as 
\be\label{indices}
 \cI (\mu_a) =
 \Tr_{\cH} (-1)^F e^{-\beta \{ Q, Q^\dagger \} } e^{-  \sum_a  \mu_a R_a} \, ,
\ee
for a choice of supercharge $Q$, and $R_a$ is the set of all R and flavor symmetries that commute with  $Q$. 
For the superconformal index \eqref{index6d}, the linear combination  $\sum_{i=1}^3 h_i + \sum_{I=1}^2 r_I$ does not commute with $Q$.
This translates into a linear constraint among the  chemical potentials 
\be\label{constraint2}
 \sum_{I=1}^{2} \Delta_I + \sum_{i=1}^{3} \omega_i =  n \, , \qquad n\in \mathbb{Z} \, .
\ee
Notice that, since $r_I$ and $J_i$ have integer eigenvalues, the chemical potentials are only defined modulo one.
For this reason the right-hand side of \eqref{constraint2} is not required to vanish but it must be an integer.
In four dimensions, where an analogous constraint appears in the definition of the  four-dimensional superconformal index,
the statement \eqref{partitionf} has been derived under the assumption $n=0$.
In six dimensions, things are less clear.

In this paper we shall consider the quantity \eqref{equivariant}, arising from the equivariant integral of the eight-form anomaly polynomial,  for a general choice
of angular ambiguities in \eqref{constraint2}. For $\fg = A_{N-1}$, the equivariant integral at large $N$ reads
\be \label{CE}
 E^{(A_{N-1})} (\Delta_I , \omega_i) = i \pi N^3 \frac{\left( \Delta_1 \Delta_2 \right)^2}{12 \omega_1 \omega_2 \omega_3} \, .
\ee

\subsection{Reproducing the entropy}
\label{entropyderivation}

We now show that the Bekenstein-Hawking entropy of the BPS black holes \eqref{7d:bh:CGLP:metric} and \eqref{7d:bh:Chow:metric}
can be obtained by extremizing the quantity
\be
 \label{extr}
 \cS (\Delta_I , \omega_i) \equiv - E^{(A_{N-1})} (\Delta_I , \omega_i)
 - 2 \pi i \sum_{I=1}^{2} \Delta_I Q_I - 2 \pi i \sum_{i=1}^{3} \omega_i J_i \, ,
\ee
where $E^{(A_{N-1})}$ is given in \eqref{CE}, with respect to $\Delta_I$, $\omega_i$ and subject to the constraint
\be
 \sum_{I=1}^{2} \Delta_I + \sum_{i=1}^{3} \omega_i = 1 \, .
\ee
In order to check it, it is convenient to work with the following parameterization of the chemical potentials:
\bea
 & \omega_1 = \frac{1}{1 + z_1 + z_2 + z_3 + z_4} \, , \qquad && \Delta_1 = \frac{z_1}{1 + z_1 + z_2 + z_3 + z_4} \, , \\
 & \Delta_2 = \frac{z_2}{1 + z_1 + z_2 + z_3 + z_4} \, , \qquad && \omega_2 = \frac{z_3}{1 + z_1 + z_2 + z_3 + z_4} \, , \\
 & \omega_3 = \frac{z_4}{1 + z_1 + z_2 + z_3 + z_4} \, .
\eea
Note that $\sum_{I=1}^{2} \Delta_I + \sum_{i=1}^{3} \omega_i = 1$.
Then the extremization equations become
\bea
 \frac{2^7 g^5 G_{\text{N}}}{\pi^2} \left( q_a - q_0 \right) & = - \frac{z_1^2 z_2^2}{z_3 z_4} ( 1 + 2 / z_a ) \, ,
 \quad \text{ for } a = 1,2 \, , \\
 \frac{2^7 g^5 G_{\text{N}}}{\pi^2} \left( q_b - q_0 \right) & = - \frac{z_1^2 z_2^2}{z_3 z_4} (1 - 1 / z_b) \, ,
 \quad \text{ for } b = 3,4 \, ,
\eea
where we have relabeled the black hole charges as
\be
 J_1 = q_0 \, , \qquad Q_1 = q_1 \, , \qquad Q_2 = q_2 \, , \qquad J_2 = q_3 \, , \qquad J_3 = q_4 \, .
\ee
The value of $\cS$ at the critical point $\tilde z_i$, as a function of the charges, is given by
\be
 \cS (Q_I,J_i) = \frac{i \pi^3}{2^6 g^5 G_{\text{N}}} \frac{\tilde z_1^2 \tilde z_2^2}{\tilde z_3 \tilde z_4} - 2 \pi i J_1 \, .
\ee
By an explicit computation one can check that the solution to the extremization equations is complex; however,
quite remarkably, $\cS$ at the critical point is a \emph{real} function of the black hole charges.
Moreover, by equating two electric charges or three angular momenta, one can check that it precisely coincides
with the entropy of the black holes \eqref{7d:bh:CGLP:metric} and \eqref{7d:bh:Chow:metric}
\be
 \label{eq:mainresult}
 \cS\big|_{\text{crit}} (Q_I , J_i) = S_{\text{BH}} (Q_I , J_i) \, .
\ee

In order to compare  the field theory inspired result  with the gravity ones in \eqref{eq:mainresult} we made use of
the relation between field theory and gravitational parameters in the large $N$ limit, which is given by
\be
 N^3 = \frac{3 \pi^2}{16 g^5 G_{\text{N}}} \, .
\ee
It is quite remarkable that the entropy of the black holes is reproduced as a Legendre transform
of the integrated anomaly polynomial with the correct field theory normalization.

\section{Discussion and conclusions}\label{sec:discussion} 

In this note we have extended our previous observation \cite{Hosseini:2017mds} that the entropy of BPS,
rotating AdS$_5\times S^5$ black holes can be written as the Legendre transform of a combination of
anomaly coefficients for R and flavor symmetries of the dual theory to the case of  AdS$_7\times S^4$
black holes. 
It would be interesting to see if the same results hold only for maximally supersymmetric dual theories or it can be also extended
to rotating black holes asymptotic to AdS$_5 \times Y_5$, where $Y_5$ is a five-dimensional Sasaki-Einstein manifold. For such black holes
there is a natural conjecture that we discuss in Appendix \ref{app:casimir}.

An important r\^ole in our analysis is played by the angular ambiguities \eqref{constraint2} in the definition of chemical potentials,
which affect both the partition function on $S^5\times S^1$ and the index. A choice of determination for the chemical potentials should be
made when performing limits, for example low- and high-temperature, or modular transformations of the integrand
of the corresponding matrix models, since these operations typically involve multi-valued functions.
Examples in the analogous four-dimensional case can be found in \cite{Assel:2014paa,Lorenzen:2014pna,Ardehali:2015hya,Ardehali:2015bla,Brunner:2016nyk}.
It is then interesting to ask whether there exists a limit in the fugacities, subject to the constraint \eqref{constraint6d},
where the quantity \eqref{6dCasimirenergy} dominates the partition function or the index.
We notice that, also for static magnetically charged black holes in AdS$_4$,
the ambiguities played a crucial r\^ole. It was shown indeed in \cite{Benini:2015eyy,Hosseini:2016tor,Hosseini:2016ume} that,
if we assume that all the real parts of chemical potentials live in the interval $[0,2\pi]$, one can find a consistent
saddle point for the topologically twisted index and reproduce the entropy of the black holes, only if the
sum of all chemical potentials is a very specific multiple of $2\pi$.%
\footnote{More precisely, with the determination $\re \Delta_I\in[0,2\pi]$, one finds a saddle point, up to discrete symmetries,
only if the sum of all $\Delta_I$ appearing in each superpotential term is $2\pi$.
The interval $[0,2\pi]$ in \cite{Benini:2015eyy,Hosseini:2016tor,Hosseini:2016ume} is analogous to the interval $[0,1]$ in this paper.} 

It would be also very interesting to compute, using supersymmetric holographic renormalization, the on-shell action of
the black holes in AdS$_5\times S^5$ and AdS$_7\times S^4$. It has been shown in \cite{Halmagyi:2017hmw,Cabo-Bizet:2017xdr,Hristov:2018lod} that,
for a class of BPS static AdS$_4$ black holes, the on-shell action indeed reproduces the entropy of the black holes and,
in the grand canonical picture, the large $N$ limit of the twisted index. It would be interesting to see if we can reproduce \eqref{6dCasimirenergy}
via a holographic computation. Notice also that anomalies seem to affect the field theory and holographic computation \cite{Papadimitriou:2017kzw,An:2017ihs}.
These anomalies could also be responsible for the choice of determination \eqref{constraint6d}.

Finally, we noticed in \cite{Hosseini:2017mds} that the extremization for AdS$_5\times S^5$ black holes with equal rotations has a nice
interpretation in terms of an attractor mechanism for static black holes in four-dimensional gauged supergravity upon dimensional reduction
of the squashed $S^3$ horizon geometry along the Hopf fiber. It would be interesting to show that a similar mechanism is at work here,
using $\cN = (1,1)$ six-dimensional gauged supergravity for the solution that is obtained by dimensional reduction of
the squashed $S^5$ horizon geometry of the black hole with equal angular momenta along the Hopf fiber. We briefly discuss the physical interpretation
of this reduction in Appendix \ref{app:reduction}.

\section*{Acknowledgements}

We would like to thank Francesco Benini and Paolo Milan for useful discussions and especially
Achilleas Passias for numerous comments and collaboration on a related project.
The work of SMH was supported by World Premier International Research Center Initiative (WPI Initiative), MEXT, Japan and in part by the INFN.
KH is supported in part by the Bulgarian NSF grant DN08/3.
AZ is partially supported by the INFN and ERC-STG grant 637844-HBQFTNCER.
SMH would like to thank the Bulgarian Academy of Sciences in Sofia and the String Theory group at University of Padova for their kind hospitality during his visit, where part of this work was done.

\appendix

\section{AdS$_5$ black holes entropy and anomalies in four dimensions}
\label{app:casimir}

In this appendix we make some  remarks on a possible generalization of the  relation between the entropy of  BPS, rotating AdS$_5$ black holes and the anomaly polynomials of their field theory duals originally presented in \cite{Hosseini:2017mds}. In particular, we consider supersymmetric black hole solutions asymptotic to AdS$_5 \times Y_5$, where the internal space $Y_5$ is a Sasaki-Einstein manifold. 

Consider five-dimensional gauged supergravity with $n_\text{V}$ massless vector multiplets and Fayet-Iliopoulos (FI) gauging. The Lagrangian is completely determined by the symmetric coefficients $C_{IJK}, I,J,K \in \{1,...,n_\text{V} \}$ which can be read off from the Chern-Simons terms in the Lagrangian,
\begin{equation}
 {\cal L}_5 = e R_{5} + \ldots - \frac16 C_{IJK} F^I \wedge F^J \wedge A^K + \ldots \, , 
\end{equation}
and the FI parameters $\xi_I$, that specify the linear combination $\xi_I A^I$ used for electrically gauging the R-symmetry. Here, $A^I$ are the $\U(1)$ gauge fields and $F^I$ their corresponding field strengths. See for example \cite{Klemm:2000gh} and references therein for a comprehensive description of five-dimensional gauged supergravity.

The general supersymmetric rotating black holes in the above class of five-dimensional gauged supergravity were written down in \cite{Kunduri:2006ek} after the seminal paper \cite{Gutowski:2004ez} and further developments. The black hole solutions, apart from explicitly depending on the numbers $C_{IJK}$ and $\xi_I$, depend on the set of asymptotic charges given by $n_\text{V}$ electric charges $Q_I$ and two angular momenta $J_{\pm}$. Due to the requirement of supersymmetry and the existence of a smooth black hole horizon, there is one additional constraint among the set of asymptotic charges. It is particularly useful to consider a Scherk-Schwarz dimensional reduction down to four dimensions as it was done in \cite{Hosseini:2017mds}, where one can explicitly write down the black hole attractor mechanism. The resulting four-dimensional supergravity has $(n_{\text{V}}+1)$ $\U(1)$ vector fields (the new Kaluza-Klein gauge field is labeled by $A^0$) and is uniquely specified by a the holomorphic prepotential,
\begin{equation}
 {\cal F} (X^\Lambda) = - \frac16 \frac{C_{IJK} X^I X^J X^K}{X^0} \, .
\end{equation}
The prepotential uniquely determines the scalar manifold given by the holomorphic sections $X^{\Lambda}, \Lambda \in \{0, I\}$, in turn defining all kinetic terms in the four-dimensional Lagrangian. The R-symmetry in four dimensions is again gauged by the linear combination $\xi_{\Lambda} A^{\Lambda}$ where the new gauge field $A^0$ is included with a weight $\xi_0 = 1$. From a four-dimensional perspective the same black holes can be described by $(n_{\text{V}} + 1)$ electric charges $(q_0 , q_I) = G^{(5)}_{\text{N}} (J_+ / 2 , - Q_I) / \pi $, an angular momentum $j = G^{(5)}_{\text{N}} J_- / 2 \pi$, and the KK magnetic charge $p^0 = 1$. Note that $G^{(5)}_{\text{N}} = 4 \pi G^{(4)}_{\text{N}}$ and the black hole entropy remains the same upon reduction to four dimensions. The static limit $(J_- = 0)$ is particularly useful since we can write down the black hole entropy, in terms of four-dimensional variables, in a compact form \cite{Cacciatori:2009iz,DallAgata:2010ejj}%
\footnote{Here we correct a sign mistake in (4.23), (4.24) and (4.30) in \cite{Hosseini:2017mds}.}
\begin{equation}\label{extrY5}
 {\cal S} (X^\Lambda) =
 - \frac{i \pi}{2 G^{(4)}_{\text{N}}} \left( q_{\Lambda} X^{\Lambda} - p^\Lambda \frac{\partial {\cal F} (X^\Lambda)}{\partial X^\Lambda} \right)
 = - \frac{i \pi}{2 G^{(4)}_{\text{N}}} \left( q_0 X^0 + q_I X^I - \frac16 \frac{C_{I J K} X^I X^J X^K}{(X^0)^2} \right) ,
\end{equation}
under the constraint
\begin{equation}
 \label{eqapp:constraint}
 \xi_{\Lambda} X^{\Lambda} = X^0 + \xi_I X^I = 1 \, .
\end{equation}
Upon extremizing $\cS(X^\Lambda)$ as given above, one fixes the scalar fields in terms of the conserved charges and recovers the correct Bekenstein-Hawking entropy at the extremum.

Introducing an extra parameter $X^-$ (being conjugate to $J_-$), in \cite{Hosseini:2017mds} we showed that the function $\cS$ can be extended to include also the last remaining charge $J_-$ for the $stu$ model. In this case the only nonvanishing triple intersection numbers are $C_{123} = 1$ (and cyclic permutation) and $\xi_I=1$. It would be interesting to similarly generalize also the $(J_- \propto j \neq 0)$ case to arbitrary parameters $C_{I J K}$, but at the moment we are lacking proper understanding of the four-dimensional rotating attractor mechanism.

However, we can try to apply these arguments  to the case of BPS, rotating black holes in AdS$_5\times Y_5$. The five-dimensional effective theory contains $n_\text{V}$ massless vector multiplets, corresponding to the R- and global symmetries of the dual field theory. Generically the reduction  on $Y_5$  leads to other matter multiplets in five-dimensional supergravity, such as hypermultiplets and massive vector multiplets. These, however, do not carry additional $\U(1)$ gauge symmetries and we will work under the assumptions that  they  decouple in the description of the  the black hole near-horizon geometry.  With this  working assumption, we could expect that the entropy is given by the minimum of \eqref{extrY5} in the case of equal angular momenta and by its natural extension for $J_-\ne 0$. This is particularly intriguing because the coefficients  $C_{IJK}$ in a compactification on AdS$_5\times Y_5$ are proportional to  the anomaly coefficients $\Tr \cQ_I \cQ_J \cQ_K$ for the $n_\text{V}$ symmetries $\cQ_I$ associated with the gauge fields  $A^I$ in the bulk five-dimensional theory \cite{Tachikawa:2005tq,Benvenuti:2006xg}. As a consequence,  it is tempting to speculate that the entropy of a black hole with electric charges $Q_I$ and angular momenta $J_i$ should be obtained as a Legendre transform of\footnote{To compare with \eqref{extrY5}, we set $X^I=\Delta_I$, $X^0=\omega_1+\omega_2$, $X^-=\omega_1-\omega_2$ and $J^{\pm}=J_1\pm J_2$.} 
\be\label{CEY5}  E(\Delta_I,\omega_i) =  - i \pi N^2 \sum_{I,J,K=1}^{n_\text{V}}  \frac{C_{I J K}}{6} \frac{\Delta_I \Delta_J \Delta_K}{\omega_1\omega_2} \, ,\ee
with respect to $\Delta_I$ and $\omega_i$ with the constraint 
\be\label{constraintY5}
 \omega_1+\omega_2 +\sum_{I=1}^{n_\text{V}} \Delta_I=1 \, .
\ee
The expression \eqref{CEY5} is fully determined by anomalies. By setting $\Delta_I =  \left(1- \omega_1 -\omega_2 \right) \hat \Delta_I / 2$, it can be written as  
\be\label{CEY52}  E(\Delta_I,\omega_i) = \frac{4 \pi i}{27} \frac{\left(  \omega_1 + \omega_2 -1 \right)^3}{\omega_1 \omega_2} a(\hat \Delta_I) \, ,\ee
where
  \be a(\hat \Delta_I)=   \frac{9 N^2}{64}  \sum_{I,J,K=1}^{n_\text{V}} C_{I J K} \hat\Delta_I \hat\Delta_J \hat \Delta_K \, , \ee
subject to $\sum_{I=1}^{n_\text{V}} \hat\Delta_I =2$ is the trial R-charge of the conformal field theory in the large $N$ limit \cite{Tachikawa:2005tq,Benvenuti:2006xg}. 
The expression \eqref{CEY52} has a strong resemblance with the refined supersymmetric Casimir energy for the Hopf surface $\cH_{p, q} \simeq S^3 \times S^1$ in the large $N$ limit \cite{Assel:2014paa,Lorenzen:2014pna,Assel:2015nca}. Indeed it differs from it only by the $-1$ in the numerator.\footnote{See, for example, Eq.\,(C.3) in \cite{Hosseini:2017mds} with 
$\omega_i=-i |b_i|$ where $p = e^{- 2 \pi |b_1|}$, $q = e^{- 2 \pi |b_2|}$.  Recall also that in the large $N$ limit $a=c$.} \eqref{CEY5} reduces exactly to the supersymmetric Casimir energy if we impose  $\omega_1+\omega_2 +\sum_{I=1}^{n_\text{V}} \Delta_I=0$ instead of \eqref{constraintY5}, corresponding to  a different choice of angular determinations for the chemical potentials.
 
\section{Dimensional reduction and topological twist on $\bC\bP^2$}
\label{app:reduction}

For the case of asymptotically AdS$_5$ BPS rotating black holes that we considered previously in \cite{Hosseini:2017mds}, the reduction of the solutions from five to four dimensions gave us an additional physical understanding. The rotating black holes with equal angular momenta reduce to static domain-wall solutions  in four dimensions with near horizon geometry AdS$_2\times S^2$. We showed that from four-dimensional perspective supersymmetry is preserved by the $\U(1)_R$ gauge field canceling the spin connection on the internal $S^2$ manifold via a topological twist \cite{Witten:1988ze,Witten:1991zz}. Without going into so much details, now we would like to argue that a similar dimensional reduction gives analogous understanding of the asymptotically AdS$_7$ black holes from six-dimensional point of view.

We focus on the single-rotation class of solutions described in section \ref{sec:single-rotation}. We can dimensionally reduce the metric and all other fields along the $\U(1)$ fiber of $S^5$, with the remaining $\bC\bP^2$ retaining all its symmetries and obtaining a static solution. We consider the usual Kaluza-Klein (KK) ansatz for the metric,
\begin{equation}  
	{\rm d} s_7^2 = e^{\phi^{\text{KK}}} {\rm d} s_6^2 + e^{-4 \phi^{\text{KK}}} ({\rm d} \psi + A^{\text{KK}})^2\ ,
\end{equation}
where $ {\rm d} s_6^2$ is the resulting 6D line element, $\phi^{\text{KK}}$ is the KK scalar field, and $A^{\text{KK}}$ the $\U(1)$ \text{KK} vector field. We see that the resulting six-dimensional line-element has the usual time and radial directions, as well as internal space $\bC\bP^2$, whose metric we repeat again here,
\bea
 \rd s^2_{\bC\bP^2} = \rd \xi^2 + \frac14 \sin^2 \xi \left( \sigma_1^2 + \sigma_2^2 + \cos^2 \xi \, \sigma_3^2 \right) \, ,
\eea
where $\sigma_i$ $(i=1,2,3)$ are left-invariant one-forms on $\SU(2)$, satisfying $\rd \sigma_i = - \frac12 \epsilon_{ijk} \sigma_j \wedge \sigma_k$. The K\"ahler form on $\bC\bP^2$ is $J = \frac12 \rd \cB$ with $\cB = \frac12 \sin^2 \xi \sigma_3$. Comparing to the explicit solution in \eqref{7d:bh:CGLP:metric}, we see that the KK vector field has a leg along the time direction carrying an electric charge in 6D (which corresponds to the angular momentum in 7D) but also has a leg along the internal $\bC\bP^2$ manifold,
\begin{equation}
\label{KKvector}
	A^{\text{KK}}_{\bC\bP^2} = \cB = \frac12 \sin^2 \xi\ \sigma_3\ .
\end{equation}
The reduction of the remaining  fields give rise to additional electric charges for the other six-dimensional vector fields. 

The only vector field along the internal manifold is  the KK vector \eqref{KKvector}, therefore we are interested in seeing how the Killing spinor covariant derivative depends on it. 
Via a general Scherk-Schwarz ansatz for the reduction of fermions along a $\U(1)$ isometry, see e.g.\, \cite{Hristov:2014eba}, the Killing spinor covariant derivative looks like
\begin{equation}
 \label{KSder}
 D_{\mu} \epsilon = \partial_{\mu} \epsilon
 + \frac{1}{4} \left( \omega_{\mu}^{\phantom{\mu}ab} \gamma_{ab}
 + 2 g A^{\text{KK}}_{\mu} \Gamma_R \right) \epsilon + \dots \, ,
\end{equation}  
where the ellipsis denotes additional connections that will not be important below. The coupling constant $g$ is left arbitrary, the $\gamma_{ab}$ are spatial gamma matrices, and the matrix $\Gamma_R$ allows for some internal structure of the spinors as typically the $\U(1)$ KK gauge field becomes part of a bigger R-symmetry mixing the fermions.

An arbitrary four-manifold has an $\SO(4) = \SU(2)_l \times \SU(2)_r$ holonomy, but for K\"ahler manifolds such as $\bC\bP^2$ we have a further simplification and one of the two $\SU(2)$ factors, say $\SU(2)_r$ becomes $\U(1)_r$. Explicitly, in the coordinates we already introduced, the nonvanishing components of the spin connection read
\begin{align}
\begin{split}
 \omega_{\sigma_1}^{\phantom{\sigma_1}14} = \omega_{\sigma_1}^{\phantom{\sigma_1}32}
 = \frac12 \cos \xi\ , \quad \omega_{\sigma_2}^{\phantom{\sigma_2}13}
 = \omega_{\sigma_2}^{\phantom{\sigma_2}24} = \frac12 \cos \xi \, , \\
 \omega_{\sigma_3}^{\phantom{\sigma_3}12}
 = \frac12 \cos^2 \xi -1\ , \quad \omega_{\sigma_3}^{\phantom{\sigma_3}34}
 = \frac12 ( \cos^2 \xi - \sin^2 \xi ) \, .
\end{split}
\end{align}
It is easy to see that splitting the spin connection into a self-dual and antiself-dual part $\omega^{\pm}$ is equivalent to splitting it into a $\U(1)_r$ and an $\SU(2)_l$ factor, respectively. In order for the supersymmetric twist to be performed, we need to cancel completely both $\omega^+$ and $\omega^-$ in the Killing spinor covariant derivative, \eqref{KSder}. The $\SU(2)_l$ part drops out automatically if we impose the projection
\begin{equation}
 \gamma^{1234} \epsilon = - \epsilon\  \quad \Rightarrow
 \quad \omega_{\mu}^{-ab} \gamma_{ab}^- \epsilon = 0 \, ,   
\end{equation}
where $2 \gamma_{ab}^- \equiv \gamma_{ab} - \varepsilon_{abcd} \gamma^{cd}$. The $\U(1)_r$ part of the spin connection is then
\begin{equation}
	\omega^{+12} = - \frac32 \sin^2 \xi\ \sigma_3 = -3 \cB = -3 A^{\text{KK}}_{\bC\bP^2} \, .
\end{equation} 
We therefore see that the spin connection is precisely canceled and the supersymmetric twist is completed upon imposing 
\begin{equation}
   \gamma^{12} \epsilon = \Gamma_R \epsilon\ , \qquad g = \frac32 \, ,
\end{equation}
in \eqref{KSder}.

Note that supersymmetric flows in six dimensions with a $\U(1)$ twist on $\bC\bP^2$ have already been explicitly found in \cite{Naka:2002jz} and further studied in \cite{Bobev:2017uzs}, in the absence of electric charges, two-form field and additional scalars. Here we have shown that the 7D rotating black holes we consider, upon reduction to six dimensions, fit in the same category of solutions in \cite{Naka:2002jz,Bobev:2017uzs} with additional conserved charges.  

\bibliographystyle{ytphys}

\bibliography{AdS7_BH}

\end{document}